A first-order phase-transition, a super-cooled fluid, and a glass in a two-dimensional lattice gas model

E. Eisenberg[1] and A. Baram[2]

1. School of Physics and Astronomy, Raymond and Beverly Sackler Faculty of Exact Sciences, Tel Aviv University, Tel Aviv 69978, Israel

2. Soreq NRC, Yavne 81800, Israel

**Abstract**. Studying the series expansion of the thermodynamic potential for the hard–core $N_3$ lattice gas model, we provide evidence for a first order phase transition with a finite jump in density and entropy, in agreement with numerical transfer matrix calculations. The solid branch terminates at the critical activity $z_c$, but the fluid branch continues beyond $z_c$, describing a super-cooled fluid. It terminates with $\rho/\rho_{cp} \sim 0.85$ and finite entropy per lattice site. This termination density is close to the random-closest-packing density of the glassy state obtained for infinitely-fast cooling. The model thus exhibits a thermodynamic meta-stable glassy phase with finite Edwards' compactivity.

The lattice gas model with nearest-neighbors hard core exclusion is long known to undergo a continuous phase transition from a low density disordered fluid state to an ordered solid phase [1-6]. On the other hand studies of the continuum model of three dimensional hard spheres (which are the limit of lattice models with an extended exclusion) indicate a first order transition with coexistence of solid and super-cooled fluid phases [7,8]. The nature of the transition in the hard disks case is not clear yet [9]. It is therefore expected that the nature of the phase transition depends on the range of the hard core interaction.

Bellemans and Orban [10], and later Orban and Van Belle [11], have studied numerically the lattice gas model of hard particles on the two dimensional square lattice, with interaction range that extends up to the third shell of nearest neighbors ($N_3$ model). This model is identical to the model of hard-core symmetric cross-shaped pentamers on the square lattice. Using the matrix method of Kramers and Wannier [3] they studied rings of infinite length and finite width M, for M=5, 10, 15 [10], and 20 [11]. The symmetry of the model requires that the width M is a multiple of 5, otherwise the system cannot reach the configuration of closest packing, whose density is $\rho_{cp} = 1/5$. Analysis of the numerical results suggests that the $N_3$ model exhibits a first order phase transition. This model is believed to be the simplest model that exhibits a first order transition. Furthermore, we have recently shown that this is also the simplest model that exhibits a glass transition, resulting from a random filling process controlled by diffusion (RSAD) [12]. This process terminates in a stable random packing state, at the density $\rho_{rcp} = 0.171626(3)$ (here and in the following, the number in parentheses is the uncertainty in the last digit). Therefore, it is worthwhile to study the properties of this exclusion model, focusing in particular on the relations between the critical values: $z_c = \exp(\beta\mu_c)$, $\rho_f(z_c)$, $\rho_s(z_c)$ (the critical activity, the density of the fluid at the transition, and that of the solid at the transition, respectively) and the termination values of the fluid and the solid branches, looking for signatures of the glassy behavior in the thermodynamic properties.

In this letter we study the nature of this phase transition through the analytic properties of the low-density and high-density expansions of the thermodynamic functions as a function of the activity. Looking at the analytic properties of these series we show the existence of a first-order transition and the coexistence of the solid phase with a super-cooled phase, extending beyond the critical activity. We support our results by extending the numerical matrix-method calculations to M=25. The termination density of the super-cooled fluid branch, as derived from extrapolating the low-density series, turns out to be very similar to $\rho_{rcp}$, the limiting density of the glassy phase obtained for infinitely fast cooling through the RSAD process [12]. This last result suggests that the super-cooled phase might have the characteristics of a glassy state for finite large activities beyond the transition. If this is the case, this model may serve as a minimal model for studying the properties of structural glasses.

We start by describing the numerical results obtained using the matrix method, which we extend here up to M=25. The approximants for the density $\rho_M(z) = \frac{z}{M}\frac{d(\ln \lambda_{max})}{dz}$ as well as their derivatives are practically M-independent outside of the critical region. However, for $0.155 \leq \rho \leq 0.19$, their M-dependence is profound. In particular, the compressibility $K = \beta \frac{d\rho}{d\mu}$ has a sharp peak, whose height diverges like $M^2$, and width decreases like $M^{-2}$. This is interpreted as a signature of a first order phase transition, where a continuous transition exhibits only a weaker, logarithmic, divergence of this peak [3]. Table 1 presents the compressibility maximum $\beta\mu_c(M)$, the pressure at $z_c = \exp(\beta\mu_c)$, the compressibility maximum and the density at the maximum for the different values of M, as well as the densities of the left (fluid) and right (solid) inflection points of the peak.

**Table 1**

| M | $\beta\mu_c$ | $\beta p(z_c)$ | $\rho(z_c(M))$ | $K_{max}$ | $\rho_f$ | $\rho_s$ |
|---|---|---|---|---|---|---|
| 10 | 3.6358 | 0.74769 | 0.1746 | .0430 | 0.1640 | 0.1860 |
| 15 | 3.6682 | 0.74408 | 0.1757 | .0879 | 0.1666 | 0.1850 |
| 20 | 3.6737 | 0.74282 | 0.17604 | .1580 | 0.1678 | 0.1844 |
| 25 | 3.6751 | 0.74225 | 0.1763 | .2629 | 0.1699 | 0.1851 |
| ∞ | 3.6762(1) | 0.74124(2) | | | | |

The critical chemical potential converges like ~ $M^{-4}$ towards it's asymptotic value, leading to $\beta\mu_c = 3.6762(1)$, or $z_c = 39.496(4)$. The pressure at the critical activity $z_c$ converges like $M^{-2}$ towards its asymptotic value $\beta p_c = 0.74124(2)$. One thus sees that the matrix method calculations provide quite accurate estimations for the critical parameters, and the nature of the global minimum of the free energy. However they do not give an accurate estimation of the density gap at the transition, and do not provide any information on the possibility of meta-stable solid or fluid phases.

We now turn to study analytically the asymptotic expansion of the thermodynamic potential. The thermodynamic functions of the solid branch can be expanded around the closest-packing configuration, in terms of u=1/z, the activity of holes. The pressure and density are given by:

$$\beta p(z) = \frac{1}{5}\{\ln(z) + \sum_{n=1}^{\infty} b_n' u^n\} \qquad (1a)$$

$$\rho(z) = \frac{1}{5}\{1 - \sum_{n=1}^{\infty} n b_n' u^n\} \qquad (1b)$$

We derive the first six coefficients $b_n'$ of the high-density series by the enumeration method of ref. [2]. The integer coefficients $nb_n'$ are presented in table 2: (the first four coefficients were given in ref [11], with a minor error in the fourth one).

**Table 2**

| $n$ | 1 | 2 | 3 | 4 | 5 | 6 |
|---|---|---|---|---|---|---|
| $nb_n'$ | 1 | 11 | 199 | 4559 | 117811 | 3287315 |

A ratio-method type analysis leads to the following unbiased fit for the coefficients:

$$nb_n' = 1.0097 \times 39.2425^{(n-1)} n^{-1.8713} \qquad (2)$$

The maximum deviation is of 1.5% for *n=2*, and the higher order coefficients are reproduced with relative accuracy better than $10^{-4}$. The value of the leading exponential term being so close to $z_c$ (as estimated from the numerical matrix method results) strongly suggest that $z_c^{-1}$ is indeed the radius of convergence of the high-density series. In addition, we verified that Levin [13] and Pade approximants of the series $n^\alpha b_n'$ for $2.80 \leq \alpha \leq 2.90$ diverge for $u \cong z_c^{-1}$. We therefore fix the value of the exponent, and fit the coefficients by the two-parameter biased expression:

$$nb_n' = 1.0388 \times 39.496^{(n-1)} n^{-1.9051} \tag{3}$$

Assuming this form holds for the higher-order coefficients as well, we may sum the series for the pressure and its derivatives at the transition point. We find the compressibility $\sim \frac{d\rho}{du}$ to diverge at $z_c$ like $(z-z_c)^{-\alpha}$ with the critical exponent $\alpha = 0.10(1)$, and the critical density and pressure are given by:

$$\rho_s(z_c) = \frac{1}{5}\{1 - \frac{1.0388}{z_c}\varsigma(1.9051) + \frac{0.0388}{z_c} - \frac{0.045}{z_c^2} + O(z_c^{-3})\} = 0.1910(1) \tag{4a}$$

$$\beta p_s(z_c) = \frac{1}{5}\{1 + \frac{1.0388}{z_c}\varsigma(2.9051) - \frac{0.0388}{z_c} + \frac{0.0225}{z_c^2} + O(z_c^{-3})\} = 0.74147(2) \tag{4b}$$

where $\varsigma(x)$ is the Riemann zeta function. Both critical parameters are in a good agreement with Levin and Pade approximants. Note that the value of the critical pressure obtained here from the series expansion is identical to the numerical value obtained by the matrix method (table 1).

The fluid phase is described by the low-density Mayer cluster expansion in powers of the activity z. The convergence radius of this series is determined by a non-physical singularity at a point on the negative z axis $z = -z_0$, which is typically very close to the origin compared to $z_c$. Using the M=25 matrix, we derived the 12$^{th}$ and 13$^{th}$ cluster coefficients (the first 11 coefficients were already found in [11]). Utilizing the 13 available cluster coefficients and the universal asymptotic form of the cluster coefficients of two dimensional repulsive models [14,15]:

$$b_n \cong (-\frac{1}{z_0})^{(n-1)} n^{-11/6}(1+cn^{-5/6}) \qquad (5)$$

we estimate the location of the leading singularity by the Levin acceleration method[13] to be $z_0 = 0.046530(1)$. The ratio $\frac{z_c}{z_0} \cong 850$ is very big and it seems, naively, that the series cannot be extrapolated to the transition region and beyond. However, one can use the well-defined pattern of the cluster coefficients of classical repulsive systems to overcome this obstacle. Taking into account the fact that the entire spectrum of the tri-diagonal matrix representations (Yang-Lee zeroes) of such systems lies on the real axis [16], it is possible to transform the series to tractable forms in spite of the screening effect of the leading singularity. The result of this approach were also confirmed by Levin summation of the activity series applying the transformation $z = w\exp(aw)$.

**Table 3**

| $n$ | 1 | 2 | 3 | 4 | 5 | 6 |
|---|---|---|---|---|---|---|
| $R_{nn}$ | 13. | 10.777778 | 10.777971 | 10.762752 | 10.755267 | 10.751491 |
| $R_{nn+1}$ | 6. | 5.495509 | 5.424623 | 5.402505 | 5.392250 | 5.386690 |

Following the matrix representation notation of [16], table 3 presents the values of the matrix elements of the symmetric tri-diagonal matrix R, defined by: $(R^n)_{11} = (n+1)|b_{n+1}|$ , where $b_n$ are the Mayer expansion coefficients. The matrix elements are obtained from the first 13 cluster coefficients. They rapidly converge to constant asymptotic values along the main and sub-diagonal, as expected for repulsive models [16]. Thus, the matrix R becomes a tri-diagonal Toeplitz-like matrix, whose asymptotic constant values B (diagonal) and A (off diagonal) are related to the two branch points of the fluid thermodynamic functions at $-z_0$ and $z_t$ by:

$$z_0^{-1} = 2A + B \qquad z_t^{-1} = 2A - B \tag{6}$$

where $z_t$ is the termination point of the fluid phase. Extrapolating the values given in table 3, one finds (see figure 2) that *2A+B=21.4907(3),* in excellent agreement with the above direct estimate of the non-physical singularity location $z_0^{-1} = 21.4915(5)$, while *2A-B* extrapolates to 0.0054(3), leading to $z_t = 180(25)$. It thus follows that one can certainly exclude the possibility of the fluid branch terminating at the transition activity, leading to the conclusion that a super-cooled fluid exists for the activity range $z_c \leq z \leq z_t$.

Using the asymptotic Toeplitz form for the R matrix, and employing the extrapolated values of the matrix elements, we computed an approximant for the density $\rho_T(z)$. The pressure of the fluid phase at the critical activity is estimated to be $\beta p_T(z_c) = 0.74083(4)$, in agreement with the matrix method estimate $\beta p_c = 0.74124(2)$. The difference is greater than the combined uncertainties, but still very small. The density at the critical activity is estimated to be $\rho_T(z_c) = 0.15803(10)$. Thus, the density gap at the transition is $\Delta\rho = 0.0330(2)$ (or 0.165(1) in units of the closest-packing density), and the entropy gap at the transition is $\frac{\Delta s}{k} = \ln(z_c)\Delta\rho = 0.1213(7)$ (per site). At the termination point of the super-cooled fluid, the density is $\rho_T(z_t) = 0.175(1)$ and the entropy is $\frac{s}{k} = 0.08(1)$. Notably, this termination density is remarkably close to the random closest packing density obtained for this model upon infinitely fast cooling [12]. It is thus tempting to suggest that as the activity increases, the super-cooled fluid branch freezes into a glass, with limiting density being the random closest packing density, and finite Edwards' compactivity [17].

In summary, we have studied the analytic properties of the high-density and low-density activity series of the thermodynamic functions for the $N_3$ lattice-gas model. We find that the solid branch terminates at finite activity, suggesting the existence of a phase transition. The termination activity and pressure are in excellent agreement with those determined by the numerical transfer matrix method, as well as with the pressure as calculated from the fluid branch at the critical activity, supporting the validity of the analytic expansion. The fluid branch does not terminate at the critical activity but rather extrapolates to describe a meta-stable super-cooled fluid phase, which terminates with finite entropy per site (Edwards' compactivity [17]) and density very close to that of a glass obtained by infinitely fast cooling of the system. We thus suggest that the simple $N_3$ lattice gas model can serve as a minimal model for the study of structural glasses.

# Figures

**Figure 1**: The density ρ (relative to the closest packing density $\rho_{cp}=1/5$) as a function of the activity z. The thick lines describe the thermodynamically stable fluid and solid phases, with a finite jump in density for the critical activity $z_c = 39.496(4)$, and the long-dashed line describes the meta-stable super-cooled fluid. Matrix method results for M=20 and M=25 are shown as well, interpolating between the fluid and the solid. The inset presents the compressibility data as calculated by the matrix method, for M=20 (dashed) and 25 (solid), exhibiting a peak at the critical activity.

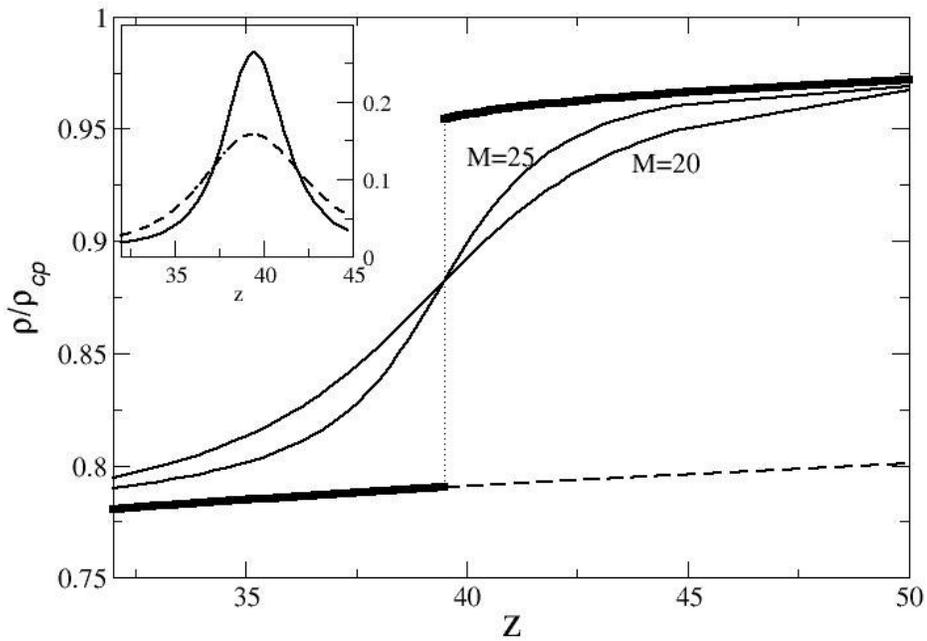

**Figure 2**: The diagonal $R_{nn}$ (circles) and twice the off-diagonal $2R_{n,n+1}$ (squares) elements of the tri-diagonal matrix R as a function of n, and their extrapolation for n>>1, as a function of $1/n^2$.

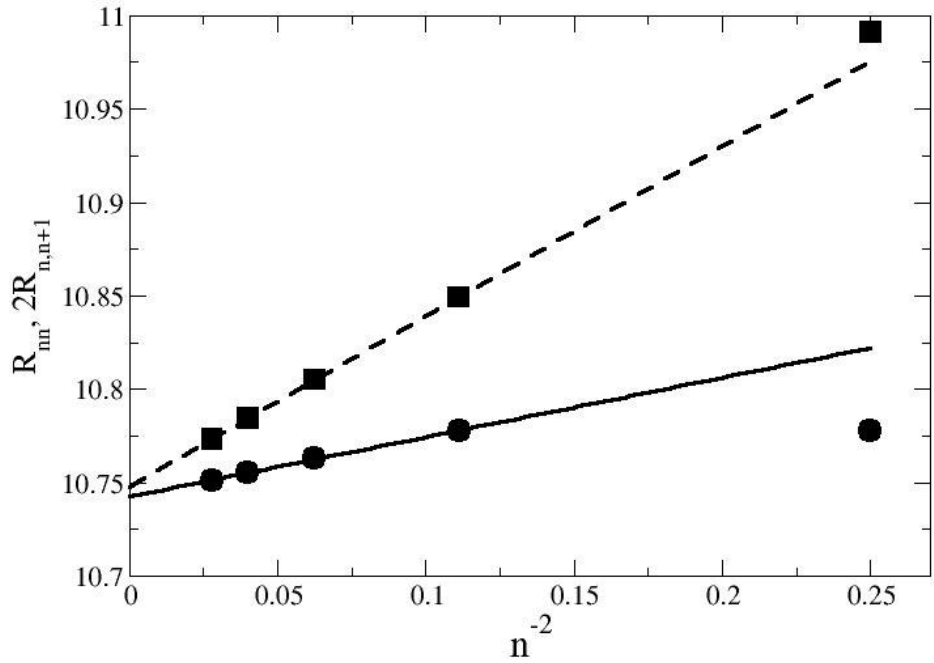